Multi-Physics Inverse Design of Varifocal Optical Devices using Data-Driven Surrogates and Differential Modeling

Zeqing Jin[1,2†], Zhaocheng Liu[1†], Nagi Elabbasi[1], Zachary Ulissi[1], Grace X. Gu[2]*, and Zhaoyu Nie[1]*

[1]Reality Labs, Meta Platforms, Inc.

[2]Department of Mechanical Engineering, University of California Berkeley

**Abstract**

Designing a new varifocal architecture in AR glasses poses significant challenges due to the complex interplay of multiple physics disciplines, including innovated piezo-electric material, solid mechanics, electrostatics, and optics. Traditional design methods, which treat each physics separately, are insufficient for this problem as they fail to establish the intricate relationships among design parameters in such a large and sensitive space, leading to suboptimal solutions. To address this challenge, we propose a novel design pipeline, mPhDBBs (multi-Physics Differential Building Blocks), that integrates these diverse physics through a graph neural network-based surrogate model and a differentiable ray tracing model. A hybrid optimization method combining evolutionary and gradient approaches is employed to efficiently determine superior design variables that achieve desired optical objectives, such as focal length and focusing quality. Our results demonstrate the effectiveness of mPhDBBs, achieving high accuracy with minimal training data and computational resources, resulting in a speedup of at least 1000 times compared to non-gradient-based methods. This work offers a promising paradigm shift in product design, enabling rapid and accurate optimization of complex multi-physics systems, and demonstrates its adaptability to other inverse design problems.

**Significance Statement**

The design of a varifocal augmented reality component is studied using an innovative artificial intelligence (AI)-driven approach. The mPhDBBs (multi-Physics Differential Building Blocks) framework is established to reveal the analytical relationships between input design variables (applied voltage and boundary support stiffness) and output objectives (focal length and focusing quality) via GNN-based surrogates and differential ray tracing model. The developed framework leverages mesh-based data structures and differentiable physics models, enabling seamless gradient flow among multiple physics. A hybrid design optimization task, utilizing both evolutionary and gradient approaches, is conducted to identify superior design candidates. This method achieves the desired multi-objectives with a speed-up of 1000 times compared to traditional non-gradient-based approaches. The developed mPhDBBs framework and hybrid optimization method can serve as a paradigm shift for computational design problems that involves large-scale design space with multi-physics simulations.

**Introduction**

Varifocal optical components have revolutionized high-quality imaging in various applications, including cameras, microscopy, and surveillance devices (1-3). With the advancement of material development and intricate structures in microelectromechanical systems, compact varifocal components have been realized in smartphones and glasses for everyday use (4, 5). The realization of adjusting focal length in these compact devices is typically driven by electrowetting for liquid lenses (6) or changing the electric field for piezoelectric materials (7). The growing demand for augmented reality (AR) and virtual reality (VR) technologies has further highlighted the importance of varifocal devices, which offer a comfortable wearing experience and reduce eye strain by enabling users to interact naturally with objects at varying distances. However, designing varifocal architectures poses significant challenges due to the complex interplay of multi-physics processes involved. The relationships between input design variables and output optical performance are difficult to explore, given the large scale and sensitive nature of these systems (8-10). Moreover, environmental factors such as temperature, humidity, and pressure can impact the performance of varifocal lenses, adding an additional layer of unpredictability to their operation. Achieving real-time, adaptive focusing performance with minimal power consumption remains a daunting task, as it requires accurate fine-tuning of the optical system, which demands high computational and energy resources. Furthermore, striking a balance between optical performance, cost-effectiveness, and durability is a delicate task, particularly for consumer devices that need to be both high-performance and affordable.

The challenge of identifying optimal design variables for a desired figure of merit (FOM) in a complex multi-physics system with a large design space is typically framed as an inverse design problem (11). In the development of optics/photonics devices, numerous design variables must be determined to produce target scattering and polarization outcomes. The relationships between FOM and design variables are often unknown and extremely complicated, making it difficult to optimize designs using only a forward model. Conventional approaches that use trial-and-error and design of experiments methods to find improved design candidates are often time-consuming and largely require hand-crafted efforts as well as prior experience by domain experts. Stochastic optimization methods, such as Gaussian process and genetic algorithms, have been used to optimize designs in a more systematic way (12-14). Recent advances in artificial intelligence have led to the development of data-driven approaches that use state-of-the-art (SOTA) forward surrogate solvers and models to simulate complex physics. These approaches enable differentiable capability during the optimization process, allowing for more efficient search for satisfactory designs. Gradient-based optimization methods, such as gradient descent and Broyden–Fletcher–Goldfarb–Shanno (BFGS) algorithms, have also been leveraged to perform design optimization (15-17). However, most design optimization studies are limited to single physics settings, such as solid mechanics or fluid dynamics, and do not account for the interactions between multiple physics. Additionally, the number of design variables is often limited to simplify the design space, resulting in an interpolation problem with a well-derived surrogate model. As a result, there is still a need for a tool that can fully reveal the analytical relationship between large-scale design space and downstream multi-physics objectives, and perform corresponding design optimization.

In this work, a framework called mPhDBBs (multi-physics differential building blocks) is developed to facilitate the opto-mechanical design of a novel AR varifocal display architecture by investigating the relationship between design variables and figure of merit (FOM) for a large-scale design space involving multi-physics simulations. The mPhDBBs framework represents a gradient-embedded modeling approach, where differentiable physics models are connected by gradient flow. This framework is then coupled with a hybrid optimization method that integrates evolutionary strategies with gradient-based techniques to identify optimal design variables that meet the desired objectives. The flow diagram of the mPhDBBs framework along with the hybrid design optimization loop is shown in Fig. 1(a). The proposed hybrid optimization approach effectively minimizes the extrapolation error by proposing new design candidates in the optimization process, realizing a balance of exploration and exploitation. Our method is validated by demonstrating its ability to reach a desired focal length with a root mean square (RMS) error within tolerance using only 2 epochs of hybrid optimization loops and 80 training samples from numerical simulation. It is believed that mPhDBBs can serve as a common paradigm and core part to realize accurate and efficient large-scale, multi-physics design optimizations in complex virtual prototyping.

**Results**

**Varifocal Optical Devices and Multi-physics Differential Building Blocks (mPhDBBs)**

In this study, we conceptualize and computationally design an AR varifocal display. The device's varifocal functionality is achieved through a deformable notch mirror that adjusts virtual image distances. An illustration of the device is provided in Fig. 1(b). The bending extent of this eyepiece-shaped mirror is designed using piezoelectric materials with applied voltages and discretized boundary support stiffness as its design parameters, as depicted in Fig. 1(c, d). The range of the virtual image distance is from infinity (no bending) to a user defined finite distance. In Fig. 1(d), optical rays are shown to illustrate the ray tracing model analysis within the region of interest, marked by a white dashed circle. Details of the deformable mirror FEA simulation setup and the ray tracing model are provided in the Materials and Methods section.

As mentioned in the Introduction section, mPhDBBs represent multi-physics, and gradient-embedded modeling, where building blocks of various differentiable physics models are connected by gradient flow. Here, the first differentiable physics model, which has solid mechanics and electrostatics embedded, is achieved via the training of GNN surrogates. The GNN-based surrogate is shown in Fig. 2 utilizing MeshGraphNet as the backbone architecture and is improved with edge augmentation methods (18-20). Nodes and meshes on the first layer (top surface) are used as input information and corresponding deformation is used as output ground truth. Edge augmentation techniques are implemented as a pre-processing step to provide a shorter path during message-passing communication and more attention in the higher deformation zone. The augmented edges can effectively mitigate the information loss during the message passing of boundary nodes towards central nodes. To create extra connections, nodes within 10 mm of the equivalent center are connected to the selected 16 nodes on the boundary perimeter. These selected nodes are the initial partition points that create the eyepiece design that is distributed uniformly on the perimeter. Both the node and augmented edge information are fed into the GNN model for encoding (feature lifting), message passing, and finally decoding to the node-level deformation profile. The subsequently connected second differentiable physics model is derived from the ray tracing algorithm with a more detailed derivation in the Materials and Methods section. The ray tracing model takes the analytical equation (fitted using Zernike terms based on the output of GNN surrogates) of the deformed mirror, the information of the rays (source and direction), and the virtual image distance as model inputs to calculate the ray intersection with the projection plane through the propagation of rays. The intersection can be further visualized using a scattered spot diagram and the spot size is used as the loss function to be minimized during optimization.

The deformation profile of the eyepiece-shaped piezo-electric mirror is obtained from a GNN-based surrogate model, which is trained based on numerical simulation ground truth. By varying the input design variables including boundary support stiffness on the perimeter and applied voltage values on piezo-electric layers, the mirror undergoes different bending extent. The data pair of input design variables and output deformation profiles are collected as datasets for surrogate training. Specifically, the input design variable includes 102 boundary support stiffness values ($W$) and 1 voltage multiplier ($V$). Fig. 3(a) shows the prediction results in the Z-axis of the deformation profile (middle column) and comparison with the simulation ground truth (left column) using 80 training samples among 100 data. The coefficient of determination ($R^2$) value reaches over 99.9% and the error between ground truth and prediction is shown in the right column with a rescaled color bar highlighting the difference.

With the differential capability obtained from trained GNN surrogates, the deformation information is fitted and the corresponding analytical surface function is fed into the ray tracing model to evaluate the optical objectives. The feasibility of mPhDBBs, especially the differential ability is validated by conducting a design optimization task to determine the superior design variables that can achieve the desired objectives including target focal length and focusing profile. The desired focal length is set as $0.59\ m$, which is a typical distance for sitting in front of a computer screen and the focusing profile is calculated as the root mean square (RMS) of the deformed surface against its perfect spherical fitting with an acceptable tolerance of $500\ nm$ as a commercial device (21, 22). A hybrid design optimization approach leveraging both evolutionary and gradient approaches is performed to locate better design variables iteratively with fresh simulation data to retrain the GNN surrogates. The detailed procedure is discussed in the Hybrid design optimization section below. After 2 epochs of iteration and a collection of around 150 simulation data points in total, a superior set of boundary support stiffness $W^*$ and voltage multiplier $V^* = 0.48742$ is obtained with a $0.59\ m$ focal length and RMS value of $481.22\ nm$ within the tolerance. The ray tracing plot and spot diagram of this satisfactory case are shown in Fig. 3(b). Additionally, $W^*$ is normalized as $[\log_{10}(W^*) - 1.70]/3.58$ and are visualized on the perimeter showing its relationship against the eyepiece geometry (Fig 3(c)).

**Hybrid design optimization for multi-objectives**

The hybrid design optimization procedure consists of three iterative subroutines: data collection, surrogate model training, and design optimization via gradient descent. In the data collection subroutine, a central design candidate is first identified, and a group of neighboring designs is then randomized. The strategy for generating the initial data batch, along with details on surrogate model training and gradient descent, is described in the Materials and Methods section. After each iteration, the newly proposed design candidate serves as the central design for the subsequent data collection subroutine, continuing the optimization loop.

In terms of the multiple objectives including focal length and focusing quality, design variables are qualitatively studied to understand their effect on the deformation profile via numerical simulations. The voltage multiplier is the dominant factor that influences the bulging of the mirror and hence affects the focal length the most. Here, the voltage multiplier is $V = [v_i]$, where $i = 1$. By sweeping the voltage multiplier from 0.1 to 0.9 with an interval of 0.1 with a fixed set of identical boundary support stiffness $W^0$, the mirror shows an increased bulging with an elliptical deformation (Fig. 2) due to the non-circular shape of the mirror. The deformed mirror has the closest extent of deformation compared to a perfect sphere with a $0.59\ m$ focal length when the voltage multiplier is 0.5 ($\bar{V} = 0.5$). Specifically, an analytical sphere function with unknown center and radius is fitted using Equation 1 and 2 based on the deformation profile and the corresponding root mean square (RMS) value is calculated using Equation 3 to indicate the deviation of the deformation against the fitted sphere. The RMS value is treated as the focusing quality. The full derivation of sphere fitting and RMS value (Equation 1-3) is discussed in the Materials and Methods section. The voltage multiplier value is a temporal and rough estimation like coarse adjustment on a microscope. The voltage multiplier value $V$ is fixed during the optimization of boundary support stiffness to avoid huge fluctuations but further tuned like fine adjustment in the last step when optimal set of boundary support stiffness $W^*$ are determined.

The optimization of boundary support stiffness follows the three subroutines. The focusing quality of collected data batch from numerical simulation is visualized using principal component analysis (PCA) (23, 24) shown in Fig. 4 (a). By abstracting relationships among design variables, PCA facilitates efficient optimization and enables the exploration of complex design landscapes. The number in the bracket indicates the number of iteration, where 0 stands for the initialization. Numbers with underline, pointing towards a cross symbol, are the central design. Numbers with a star superscription, directing into a star symbol, are the best candidate among the neighbor designs of the current iteration. Fig. 4 (b, c) show the result of gradient descent in the surrogate model prediction domain, where the first candidate (1) is proposed based on the surrogates trained on initialization data batch and the second candidate (2) is proposed based on the surrogate model trained on all previous data. The initialization position of the gradient descent starts from the previous best candidate design (0* or 1*). The surrogate model's prediction for the candidate designs are shown in grey scale background with a darker color indicating smaller RMS values. Fig. 4 (d, e) visualize the spot diagram of the proposed set of boundary support stiffness at the virtual image distance which has the minimum RMS value. Here, 2* is the satisfactory set of boundary support stiffness $W^*$ with RMS value less than 500 nm and a focal length of 580.23 $mm$. This is exactly the same set of stiffness shown in Fig. 3 (b, c) but with a suboptimal voltage multiplier.

Lastly, the voltage multiplier is fine-tuned from the predetermined value $\bar{V} = 0.5$, based on interpolation. Specifically, several simulation data are collected using the best set of stiffness $W^*$ and voltage multiplier close to $\bar{V}$. Each simulation leads to a deformation profile with the optimal focal length and corresponding RMS value as shown in the spot diagram. A linear fitting is then performed (Focal length $= -776.63 v_1 + 968.545$) to project the focal length based on voltage multiplier input. By setting the desired focal length to 590 $mm$, the required voltage multiplier $V^*$ can be determined. A Pseudocode for the developed hybrid optimization pipeline is provided in Table. S1.

**Discussion**

A multi-physics, differential modeling framework mPhDBBs is developed to analytically reveal the relationship of a varifocal optical component between its design variables, especially boundary support stiffness with respect to multiple optical objectives. A hybrid design optimization process is developed to perform a multi-objective inverse design task, aiming at a desired focal length and focusing quality. It utilizes both evolutionary and gradient approaches to mitigate extrapolation error by involving fresh simulation data during the optimization loop. A superior set of 103 design variables is reached with 2 epochs of iteration using 150 data samples, realizing the desired focal length of 590 $mm$ with an RMS value of 481.22 $nm$ within tolerance.

Moreover, by implementing differentiable surrogate models and ray tracing physics models, the pipeline has great potential to solve more complicated problems with coupled physics and limited datasets as future works. For instance, two differentiable physics models can be connected in parallel to find an equilibrium solution for such coupled physics simulation (i.e. flutter-wind effects on structures) (25, 26). Adjoint methods and physics-informed neural networks can be derived to build differentiable physics models with limited data points by penalizing the residuals of partial differential equations (27, 28). Additionally, prior knowledge of the problem could be also included in the gradient descent procedure to better explore the superior design candidates. It is believed that such AI-driven approaches can provide huge acceleration as a paradigm shift towards the virtual prototyping of large-scale, multi-physics, and multi-objective inverse design challenges.

**Materials and Methods**

**Multi-physics simulation setup**

The simulation of the eyepiece-shaped, multi-layer piezo-electric mirror is conducted using COMSOL. It has a total of 10 layers, where the first layer is defined as the optical notch reflection layer, the second, fourth, eighth, and tenth layers are assigned as the PVDF layer, the third, fifth, seventh, and ninth layers are assigned as PMMA layer as adhesives, the sixth layer is assigned with ultra-thin glass material. Except for PVDF layers, other layers are all treated as linear elastic material. At the top and bottom surfaces of the mirror, spring dampers are applied at the nodes on the perimeter, mimicking boundary support stiffness. The top and bottom surface share the same stiffness distribution across the perimeter. The top surface has $N = 651$ nodes and $M = 3448$ edges. Electric potential values $(v_1 V_0^1, v_1 V_0^2, \ldots, v_1 V_0^8)$ are applied at the eight interfaces of PVDF layers, where $v_1$ is the voltage multiplier. The gravitational effect is applied to the whole system. Simulation is performed for quasi-static actuation by the increment of $\eta$. For a fixed voltage multiplier, the simulation takes 2 hours to collect 50 results with different boundary support stiffness designs using parametric sweep on a desktop equipped with AMD Ryzen Threadripper PRO 5975WX with 32 cores and 128 GB RAM. The displacement in three directions of all nodes is recorded in tabular sheets as simulation results.

**Dataset initialization**

Due to the large design space of over a hundred stiffness that vary from 100 to 200,000 $N/m$, random generation of stiffness values has minimal chances of achieving a deformation profile close to a perfect sphere. The sampling strategy starts from a hypothesis that the boundary support stiffness is positively correlated with the distance $d$ of its location and the equivalent center. The equivalent center of the mirror is the location that undergoes the most deformation in the first step when the voltage multiplier is 0.5. The stiffness and distances are normalized between 0 and 1 with respect to their lower and upper bounds. The boundary support node that has the smallest and largest distance is assigned with 100 (lower bound) and 200,000 (upper bound) $N/m$ respectively. For the remaining boundary support nodes, the relationship function between normalized stiffness and normalized distance is determined using a Bézier curve $\hat{G}(a, b)$, which has two fixed points at (0,0) and (1,1) as well as a controlling point moving within the unit length bounding box, generating a nonlinear curve (29). Two sets of design of experiments (DOE) are performed on the location of the controlling point with the first set defined on a coarse grid with 0.1 interval (Fig. S1(a)) and the second set focused on the zoom-in region, which has lower RMS values (Fig. S1(b)) based on the first set results. The smallest RMS value reaches 611.09 $nm$ when the controlling point is located at (0, 0.38) and this Bézier curve is used to initialize the first set of boundary support stiffness ($W^p = \underline{0}$). The first set of stiffness is then augmented to 50 sets by randomly varying each element within 20% of its value $N(W^p, \varepsilon_{[0.8,1.2],102})$ to generate the initialization data batch for training the surrogate model $\widehat{S^p}$.

**GNN surrogates and gradient descent**

After the initialization and first iteration, 100 data points are collected. The training of surrogate model uses 80% and the remaining 20% for testing. The training and evaluation of the testing dataset takes 0.5 hours for 2000 iterations with a batch size equal to 4 using a desktop equipped with a NVIDIA GeForce RTX 3080 GPU. The gradient descent method is applied due to the differentiable surrogates and ray tracing model and implemented using Pytorch with automatic differentiation and Adam optimizer. The optimization converges after 500 epochs and takes around 1 minute.

**Analytical fitting of the deformation profile**

To get the analytical equation of the deformed surface, two steps of fitting are performed using sphere fitting (Equation 1 and 2) and Zernike terms fitting (Equation 4 and 5). Assume the node information input is denoted as $X = \begin{bmatrix} X_1, Y_1, Z_1 \\ X_2, Y_2, Z_2 \\ \vdots \\ X_N, Y_N, Z_N \end{bmatrix}$ and the deformation output is denoted as $\Delta = \begin{bmatrix} dx_1, dy_1, dz_1 \\ dx_2, dy_2, dz_2 \\ \vdots \\ dx_N, dy_N, dz_N \end{bmatrix}$, the standard sphere equation can be expressed as $(x - x_0)^2 + (y - y_0)^2 + (z - z_0)^2 = r^2$. The following matrices are constructed:

$$X_{\text{current}} = X + \Delta = \begin{bmatrix} X_1, Y_1, Z_1 \\ X_2, Y_2, Z_2 \\ \vdots \\ X_N, Y_N, Z_N \end{bmatrix} + \begin{bmatrix} dx_1, dy_1, dz_1 \\ dx_2, dy_2, dz_2 \\ \vdots \\ dx_N, dy_N, dz_N \end{bmatrix} = \begin{bmatrix} x_1, y_1, z_1 \\ x_2, y_2, z_2 \\ \vdots \\ x_N, y_N, z_N \end{bmatrix} \quad (1)$$

$$A_1 = \begin{bmatrix} 2x_1, 2y_1, 2z_1, 1 \\ 2x_2, 2y_2, 2z_2, 1 \\ \vdots \\ 2x_N, 2y_N, 2z_N, 1 \end{bmatrix}, \xi_1 = \begin{bmatrix} x_0 \\ y_0 \\ z_0 \\ r^2 - x_0^2 - y_0^2 - z_0^2 \end{bmatrix}, b = \begin{bmatrix} x_1^2 + y_1^2 + z_1^2 \\ x_2^2 + y_2^2 + z_2^2 \\ \vdots \\ x_N^2 + y_N^2 + z_N^2 \end{bmatrix}$$

The unknown radius $r$ and origin $(x_0, y_0, z_0)$ can be fitted by solving $A_1 \xi_1 = b$ using least squares. The analytical equation of perfect sphere can be express as:

$$z_{\text{sphere}} = \sqrt{r^2 - (x - x_0)^2 - (y - y_0)^2} + z_0 \quad (2)$$

$$z' = \begin{bmatrix} \sqrt{r^2 - (x_1 - x_0)^2 - (y_1 - y_0)^2} + z_0 \\ \sqrt{r^2 - (x_2 - x_0)^2 - (y_2 - y_0)^2} + z_0 \\ \vdots \\ \sqrt{r^2 - (x_N - x_0)^2 - (y_N - y_0)^2} + z_0 \end{bmatrix} = \begin{bmatrix} z_1' \\ z_2' \\ \vdots \\ z_N' \end{bmatrix}, z = \begin{bmatrix} z_1 \\ z_2 \\ \vdots \\ z_N \end{bmatrix}$$

$$\text{RMS} = \sqrt{\frac{1}{N} \sum_{i=1}^{N} (z_i' - z_i)^2} \quad (3)$$

The output profile is then centered in $x$ and $y$ direction as an offset to align the ray tracing.

$$X_{\text{current}}' = \begin{bmatrix} x_1 - x_0, y_1 - y_0, z_1 \\ x_2 - x_0, y_2 - y_0, z_2 \\ \vdots \\ x_N - x_0, y_N - y_0, z_N \end{bmatrix} = \begin{bmatrix} x_1', y_1', z_1 \\ x_2', y_2', z_2 \\ \vdots \\ x_N', y_N', z_N \end{bmatrix} \quad (4)$$

Zernike terms are constructed as follows with each individual representation shown in Table. S2:

$$A_2 = \begin{bmatrix} Z_0^0(x_1',y_1'), Z_1^{-1}(x_1',y_1'), Z_1^1(x_1',y_1'), Z_2^{-2}(x_1',y_1'), Z_2^0(x_1',y_1'), Z_2^2(x_1',y_1') \\ Z_0^0(x_2',y_2'), Z_1^{-1}(x_2',y_2'), Z_1^1(x_2',y_2'), Z_2^{-2}(x_2',y_2'), Z_2^0(x_2',y_2'), Z_2^2(x_2',y_2') \\ \vdots \\ Z_0^0(x_N',y_N'), Z_1^{-1}(x_N',y_N'), Z_1^1(x_N',y_N'), Z_2^{-2}(x_N',y_N'), Z_2^0(x_N',y_N'), Z_2^2(x_N',y_N') \end{bmatrix}$$

$$\xi_2 = [\eta_0^0 \quad \eta_1^{-1} \quad \eta_1^1 \quad \eta_2^{-2} \quad \eta_2^0 \quad \eta_2^2]^T$$

The unknown coefficient attached to each Zernike term $\xi_2$ can be fitted by solving $A_2 \xi_2 = z$ using least squares. The fitted analytical equation of deformed surface can be expressed as:

$$z_{\text{Zernike}} = \eta_0^0 Z_0^0(x,y) + \eta_1^{-1} Z_1^{-1}(x,y) + \eta_1^1 Z_1^1(x,y) + \eta_2^{-2} Z_2^{-2}(x,y) + \eta_2^0 Z_2^0(x,y) + \eta_2^2 Z_2^2(x,y) \quad (5)$$

It is worth noting that the first six Zernike terms are used to fit the deformed surface for the following two reasons: 1) Focusing on dominant optical terms such as defocus ($Z_2^0$) and primary astigmatism ($Z_2^{-2}$ and $Z_2^2$) helps capture the main trends of deformation and ease of convergence during the optimization process. 2) Although the surface fitting could lose some resolution at higher order Zernike terms, it still provides a reliable direction for gradient descent in the optimization process and is able to successfully achieve the desired objectives within tolerance.

**Differentiable ray tracing model**

The ray tracing model consists of a plane source, a reflective mirror parametrized by the GNN, and a detector. The source emits rays that are perpendicular to the source plane and hit the reflective mirror. The detector records the ray locations of all the reflected rays. We formulate the loss function of the optimization as the RMS of the difference of the ray locations and the desired focal length.

Differential ray tracing serves as a fast and robust method to calculate gradients of optical system parameters. It has been investigated and researched in various subjects in optics and graphics (30-32). The fundamental idea of differential ray tracing is to utilize automatic differentiation to implement ray tracing such that the gradient of all parameters can be identified using backpropagation. The implementation of differential ray tracing varies depending on the parametrizations of the geometry and surface, and here we focus on the method of analytical surface representation: $z = f(x, y; \theta)$, where $x$ and $y$ are the spatial coordinate and $\theta$ represents the parameter vector we need to optimize. In the differential ray tracing model with Zernike polynomial, $\theta$ can be treated as the coefficients of the polynomial. When the gradient of $\theta$ is computed, we can backpropagate it to the design parameter of the mechanical parameters of the deformable mirror. The general ray tracing of the model is as follows:

**a.** We define the source with ray origins $o$ and directions $v$, where $o = [o_1, o_2, \ldots, o_N]$ and $u = [u_1, u_2, \ldots, u_N]$, assuming a total of $N$ rays are shooting from the source. The rays thus can be represented as lines: $c(t) = o + ut$, where $t$ is the time vector that the ray marches along the direction $u$.

**b.** We then find its intersection with the surface $z = f(x, y; \theta)$ using Newton's method. Notice that in Newton's method, we need to find the gradient of the surface with respect to the spatial variables $\partial f / \partial x$ and $\partial f / \partial y$. This can be easily achieved by automatic differentiation. We represent the intersection points as $q(t)$.

**c.** Given the intersection points, we can apply reflection and reflection to derive the output ray direction. The vector formulation of reflection and refraction are:

$$u_r = u - 2(u \cdot n(q)) \quad (5)$$

$$u_t = n(q)\sqrt{1 - \mu^2[1 - (n(q) \cdot u)^2]} + \mu[u - (n(q) \cdot u)n(q)] \quad (6)$$

, where $n$ is the normal vector at the intersection point and $\mu = n_{inc}/n_{trn}$ is the ratio of refractive indices of incident and transmit sides. Normal vector $n$ can be derived through automatic differentiation as: $n = [\partial f/\partial x, \partial f/\partial y, -1]^T$

**d.** Given the intersection points and output direction, we can seek to intersect the next surface and repeat the process, until the rays hit the detector. The gradients of all the procedures above can be computed using automatic differentiation. In the step of Newton's method to find intersections, if the gradient is detached for memory saving, the formulation in this study (33) can be used to find the gradient.

References

1. S. Kang, M. Duocastella, C. B. Arnold, Variable optical elements for fast focus control. *Nat. Photonics* **14**, 533-542 (2020).
2. M. J. Mescher, M. L. Vladimer, J. J. Bernstein (2002) A novel high-speed piezoelectric deformable varifocal mirror for optical applications. in *Technical Digest. MEMS 2002 IEEE International Conference. Fifteenth IEEE International Conference on Micro Electro Mechanical Systems (Cat. No.02CH37266)*, pp 511-515.
3. Y. Zhou, J. Zhang, F. Fang, Design of the varifocal and multifocal optical near-eye see-through display. *Optik* **270**, 169942 (2022).
4. C. A. Dirdal *et al.*, MEMS-tunable dielectric metasurface lens using thin-film PZT for large displacements at low voltages. *Opt. Lett.* **47**, 1049-1052 (2022).
5. A. Roszkiewicz, M. Garlińska, A. Pregowska, Advancements in Piezoelectric-Enabled Devices for Optical Communication. *physica status solidi (a)* **n/a**, 2400298.
6. X. Song, H. Zhang, D. Li, D. Jia, T. Liu, Electrowetting lens with large aperture and focal length tunability. *Scientific Reports* **10**, 16318 (2020).
7. M. C. Wapler, Ultra-fast, high-quality and highly compact varifocal lens with spherical aberration correction and low power consumption. *Opt. Express* **28**, 4973-4987 (2020).
8. A. Michael, C. Y. Kwok, Piezoelectric micro-lens actuator. *Sensors and Actuators A: Physical* **236**, 116-129 (2015).
9. S.-H. Chen, A. Michael, C. Y. Kwok, Enhancing out-of-plane stroke in piezoelectrically driven micro-lens actuator with residual stress control. *Sensors and Actuators A: Physical* **303**, 111620 (2020).
10. F. Lemke *et al.*, Multiphysics simulation of the aspherical deformation of piezo-glass membrane lenses including hysteresis, fabrication and nonlinear effects. *Smart Materials and Structures* **28**, 055024 (2019).
11. C. Kang *et al.*, Large-scale photonic inverse design: computational challenges and breakthroughs. *Nanophotonics* **13**, 3765-3792 (2024).
12. Y. Wang, Y. Lian, Z. Wang, C. Wang, D. Fang, A novel triple periodic minimal surface-like plate lattice and its data-driven optimization method for superior mechanical properties. *Applied Mathematics and Mechanics* **45**, 217-238 (2024).
13. S. Koziel, A. Pietrenko-Dabrowska, On nature-inspired design optimization of antenna structures using variable-resolution EM models. *Scientific Reports* **13**, 8373 (2023).

14. A. E. Gongora *et al.*, A Bayesian experimental autonomous researcher for mechanical design. *Science advances* **6**, eaaz1708 (2020).
15. T. Zhou *et al.*, AI-aided geometric design of anti-infection catheters. *Science Advances* **10**, eadj1741 (2024).
16. J. Reinhard, P. Urban, S. Bell, D. Carpenter, M. S. Sagoo, Automatic data-driven design and 3D printing of custom ocular prostheses. *Nat. Commun.* **15**, 1360 (2024).
17. T. Xue *et al.*, JAX-FEM: A differentiable GPU-accelerated 3D finite element solver for automatic inverse design and mechanistic data science. *Comput. Phys. Commun.* **291**, 108802 (2023).
18. T. Pfaff, M. Fortunato, A. Sanchez-Gonzalez, P. W. Battaglia, Learning mesh-based simulation with graph networks. *arXiv preprint arXiv:2010.03409* (2020).
19. Z. Jin, B. Zheng, C. Kim, G. X. Gu, Leveraging graph neural networks and neural operator techniques for high-fidelity mesh-based physics simulations. *APL Machine Learning* **1** (2023).
20. R. J. Gladstone *et al.*, Mesh-based GNN surrogates for time-independent PDEs. *Scientific reports* **14**, 3394 (2024).
21. S. Wickenhagen, S. Kokot, U. Fuchs (2017) Tolerancing aspheres based on manufacturing knowledge. in *Optifab 2017* (SPIE), pp 140-144.
22. J. Béguelin, W. Noell, T. Scharf, R. Voelkel, Tolerancing the surface form of aspheric microlenses manufactured by wafer-level optics techniques. *Appl. Opt.* **59**, 3910-3919 (2020).
23. F. Pedregosa *et al.*, Scikit-learn: Machine learning in Python. *the Journal of machine Learning research* **12**, 2825-2830 (2011).
24. K. Pearson, LIII. On lines and planes of closest fit to systems of points in space. *The London, Edinburgh, and Dublin philosophical magazine and journal of science* **2**, 559-572 (1901).
25. E. Antimirova, J. Jung, Z. Zhang, A. Machuca, G. X. Gu, Overview of computational methods to predict flutter in aircraft. *J. Appl. Mechanics.* **91** (2024).
26. I. E. Garrick, W. H. Reed III, Historical development of aircraft flutter. *Journal of Aircraft* **18**, 897-912 (1981).
27. X. Zhao, F. Mezzadri, T. Wang, X. Qian, Physics-informed neural network based topology optimization through continuous adjoint. *Structural and Multidisciplinary Optimization* **67**, 143 (2024).
28. X. Chen, L. Yang, J. Duan, G. E. Karniadakis, Solving Inverse Stochastic Problems from Discrete Particle Observations Using the Fokker--Planck Equation and Physics-Informed Neural Networks. *SIAM Journal on Scientific Computing* **43**, B811-B830 (2021).
29. D. Hermes, Helper for Bézier curves, triangles, and higher order objects. *Journal of Open Source Software* **2**, 267 (2017).
30. X. Yang, Q. Fu, W. Heidrich, Curriculum learning for ab initio deep learned refractive optics. *Nat. Commun.* **15**, 6572 (2024).
31. Z. Zhu, Z. Liu, C. Zheng, Metalens enhanced ray optics: an end-to-end wave-ray co-optimization framework. *Opt. Express* **31**, 26054-26068 (2023).
32. V. Sitzmann *et al.*, End-to-end optimization of optics and image processing for achromatic extended depth of field and super-resolution imaging. *ACM Transactions on Graphics (TOG)* **37**, 1-13 (2018).
33. L. Yariv *et al.*, Multiview neural surface reconstruction by disentangling geometry and appearance. *Advances in Neural Information Processing Systems* **33**, 2492-2502 (2020).

**Figures and Tables**

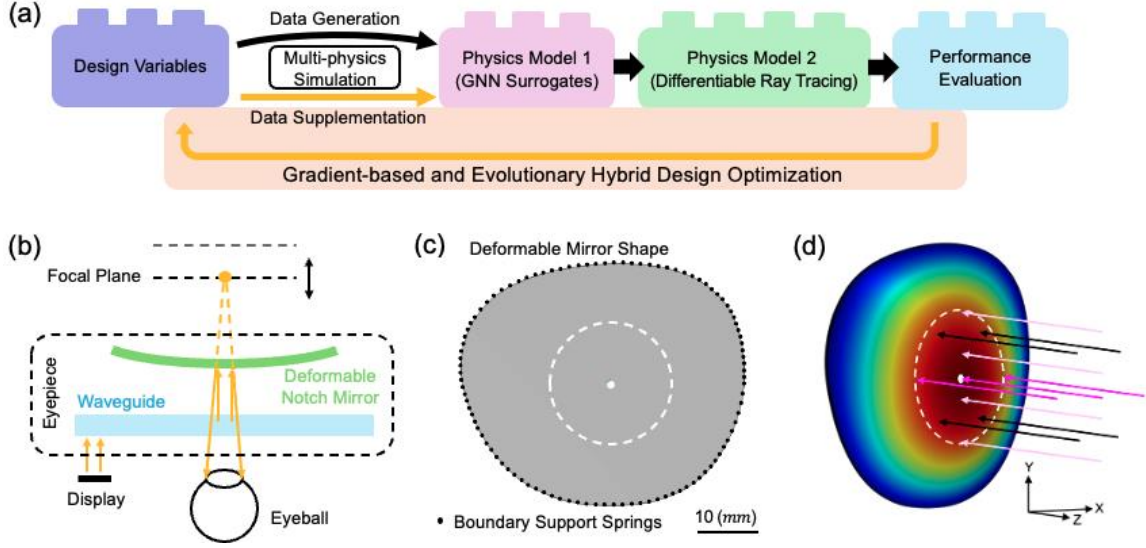

Figure 1. (a) The illustration of mPhDBBs pipeline and a hybrid design optimization process, designed to facilitate multi-physics and multi-objective optimization. mPhDBBs integrate multiple physics models through conserving gradients. (b) A schematic diagram of an AR varifocal display architecture is presented, featuring a waveguide display and a piezo-electric deformable notch mirror. The focus of this work lies in the opto-mechanical design of the deformable mirror, with the goal of achieving at least two virtual image distances. (c) The deformable mirror has a non-rotationally symmetric eye shape with discrete boundary support springs along the perimeter. The region of interest (ROI) is highlighted by a white dashed circle. (d) Illustration of incoming rays interacting with the deformable mirror in the ray tracing model.

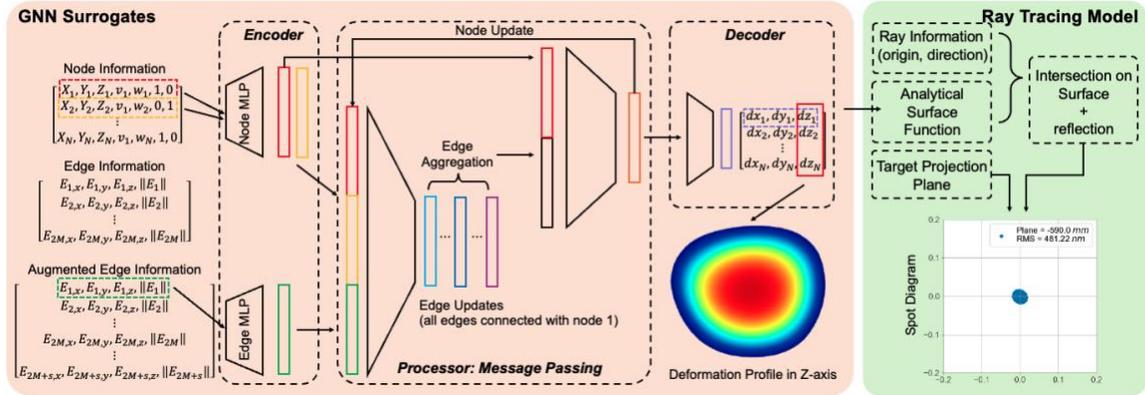

Figure 2. Illustrated is the multi-physics differential modeling framework, which integrates a data-driven GNN surrogate (pink region) with a differentiable ray tracing model (green background). The GNN surrogate, based on an edge-augmented MeshGraphNet architecture, predicts the 3D shape of the piezo-electric deformable lens given design variables: voltage multiplier $v_1$ and boundary support spring stiffness $w_i$. The predicted 3D shape and its gradient are then fed into the differentiable ray tracing model, which evaluates the system imaging quality by calculating the spot diagram and provides gradients to the design variables via backpropagation.

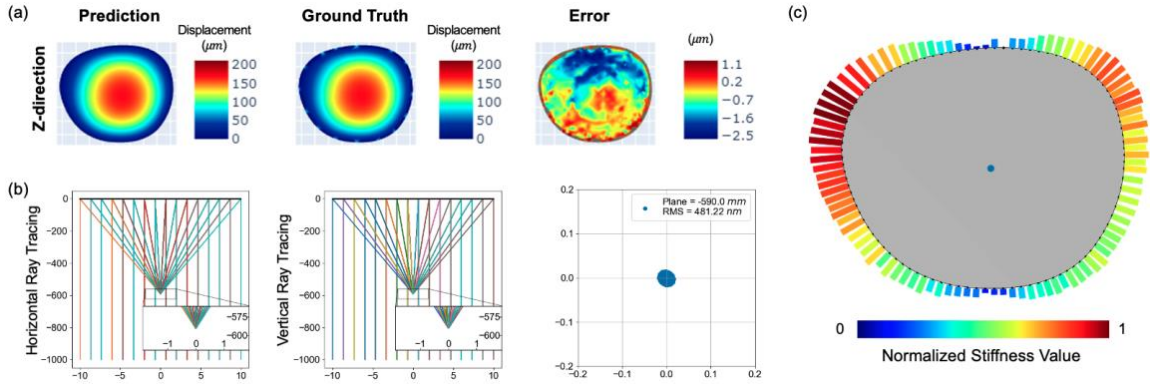

Figure 3. (a) Visualization of the mirror deformation profile. The left panel shows predictions from the GNN surrogate model, the middle panel presents FEA simulation validation, and the right panel displays the discrepancy between the predictions and the FEA simulation, using a modified color scale. (b) The left and middle subfigures show the ray tracing results along the horizontal and vertical directions. Rays travel from bottom to top, hitting the deformable mirror and reflecting to focus at the desired virtual image distance. The spot diagram shown on the right subfigure illustrates the display quality at -590 $mm$. (c) The distribution of normalized boundary support stiffness that achieve the desired focusing quality (one of the objectives).

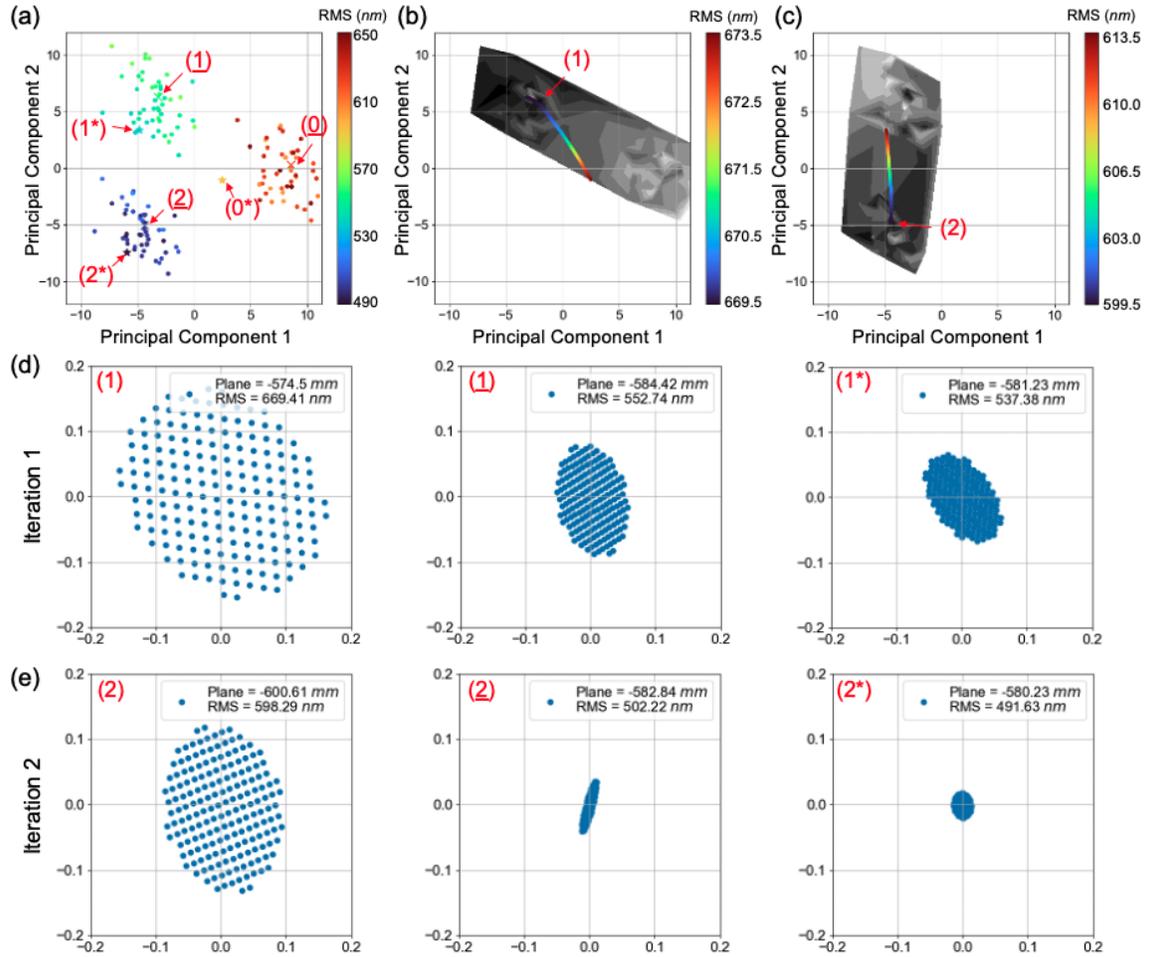

Figure 4. Illustrated is the hybrid design optimization process in mPhDBBs, which leverages principal component analysis (PCA) to navigate the high-dimensional design space of boundary support stiffness. (a) Visualization of proposed design candidates in principle component axes. RMS of the spot diagram is one of the objective functions to be minimized. Cross symbols (number with underline) indicate the central design and star symbols (number with star) are the best design candidates of each data batch (central design and neighbor designs). (b, c) The gradient descent process to find the optimized design candidates in the first and second iteration based on the surrogates. (d, e) Spot diagrams illustrate the focusing quality evolution after each subroutine: optimized design via gradient descent, central design in numerical simulation, and best design among each data batch.

**Supplementary information**

**Pseudocode for hybrid design optimization**

Here, we denote the input design variables are $\Lambda = [\lambda_1, \lambda_2, \ldots, \lambda_{i+j}]$. The desired multi-objectives are $D^* = D(\Lambda^*)$ and $S^* = S(\Lambda^*)$, where $D$ and $S$ represent an independent evaluation of design variables based on numerical simulation results and $\Lambda^*$ indicates the superior design variables meet the desired objectives. The design variables are split into two groups based on their correlation with respect to input variables. Hence, we have $\Lambda = [V, W]$, where $V = [v_1, v_2, \ldots, v_i]$, $W = [w_1, w_2, \ldots, w_j]$, and $D^* = D(\Lambda^*) = D(V^*, W^*)$ and $S^* = S(\Lambda^*) = S(V^*, W^*)$. The correlation of function $F(x, y)$ with respect to input variable $x$ is denoted as $\rho_{x,F}$. We assume $|\rho_{V,D}| \gg |\rho_{W,D}|$ and $|\rho_{W,S}| \gg |\rho_{V,S}|$. The continuous objective function obtained from surrogate model is denoted as $\widehat{D}(V, W)$ and $\widehat{S}(V, W)$. Table 1 shows the pseudocode for hybrid optimization process to find superior design variables $V^*, W^* = \underset{V,W}{\operatorname{argmin}} |D(V, W) - D^*| + |S(V, W) - S^*|$.

**Table S1. Pseudocode for multi-objective, gradient-based, and evolutionary optimization algorithm**

**Step 1:** With a fixed initialization of $W^0 = [w_1^0, w_2^0, \ldots, w_j^0]$, where $w_1^0 = w_2^0 = \ldots = w_j^0$, find $\bar{V} = \underset{V}{\operatorname{argmin}} |D(V, W^0) - D^*|$ in numerical simulation by running $m$ sets DOE.

**Step 2:** Find $W^p = \underset{W}{\operatorname{argmin}} |S(\bar{V}, W) - S^*|$, where $W = \hat{G}(a, b)$ is a user-defined hypothesis function, by running $n$ sets DOE based on variables $a$ and $b$.

**Step 3:** Surrogate model training $\widehat{S^p}$ based on dataset with $k$ sets input design variables $[\bar{V}, N(W^p, \varepsilon_{[0.8,1.2],j})]$, where $N(W^p, \varepsilon_{[0.8,1.2],j}) = W^p \odot \operatorname{unif}_{[0.8,1.2],j}$. Symbol $\odot$ represents an element-wise multiplication and $\operatorname{unif}_{[0.8,1.2],j}$ indicates a vector of uniform distribution of $j$ elements ranging from 0.8 to 1.2.

**Step 4:** Find $W^q = \underset{W}{\operatorname{argmin}} |\widehat{S^p}(\bar{V}, W) - S^*|$ using gradient descent method based on the surrogate model.

**Step 5:** Determine if the proposed set of design variables $|S(\bar{V}, W^q) - S^*| < \operatorname{tol}$ or any of its $k$ sets neighbor design variables $|S(\bar{V}, N(W^q, \varepsilon)) - S^*| < \operatorname{tol}$. If not, repeat **step 3** and **step 4** to get new surrogate model $\widehat{S^{p+1}}$ based on design variables $[\bar{V}, N(W^q, \varepsilon)]$ as well as optimized new candidate design $W^{q+1} = \underset{W}{\operatorname{argmin}} |\widehat{S^{p+1}}(\bar{V}, W) - S^*|$ until satisfying the tolerance. Assuming after $r + 1$ epochs of iteration, $W^{q+r} = \underset{W}{\operatorname{argmin}} |\widehat{S^{p+r}}(\bar{V}, W) - S^*|$ and one design set $W^*$ within $N(W^{q+r}, \varepsilon)$ meets $|S(\bar{V}, W^*) - S^*| < \operatorname{tol}$.

**Step 6:** Find $V^* = \underset{V}{\operatorname{argmin}} |D(V, W^*) - D^*|$ by interpolating linearly within the set of design variables $[N(\bar{V}, \varepsilon_{[0.95,1.05],i}), W^*]$.

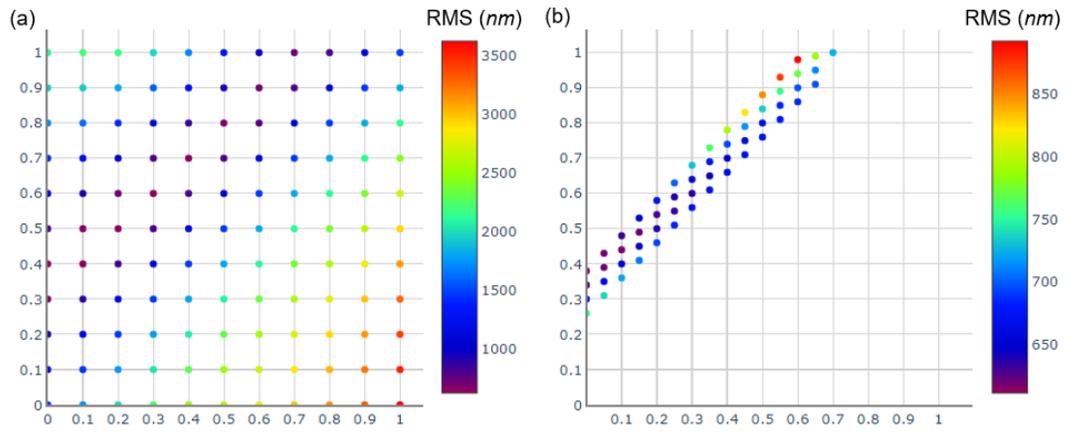

Figure S1. Design of experiments to find the initialized design variables based on user defined hypothesis.

Table S2. The first six Zernike terms in Cartesian coordinate.

| n | m | $Z_n^m$ |
|---|---|---|
| 0 | 0 | 1 |
| 1 | -1 | $2y$ |
| 1 | 1 | $2x$ |
| 2 | -2 | $2\sqrt{6}xy$ |
| 2 | 0 | $\sqrt{3}[2(x^2 + y^2) - 1]$ |
| 2 | 2 | $\sqrt{6}(x^2 - y^2)$ |